\documentclass[12pt]{elsart}
\usepackage{amstex}
\usepackage{amssymb}
\usepackage{citesort}
\usepackage{epsfig}

\begin{document}

\begin{flushright}
BARI-TH 229/96 
\end{flushright}

\begin{frontmatter}

\title{Unstable Modes and Confinement \\ in the Lattice Schr\"odinger
Functional Approach}

\author[INFN,Dep]{Paolo Cea\thanksref{emcea}} and
\author[INFN]{Leonardo  Cosmai\thanksref{emcosmai}} 
\address[INFN]{INFN - Sezione di Bari - 
Via Amendola, 173 - I 70126 Bari - Italy} 
\address[Dep]{Dipartimento di Fisica Univ. Bari -
Via Amendola, 173 - I 70126 Bari - Italy} 
\thanks[emcea]{E-mail: cea@@bari.infn.it}
\thanks[emcosmai]{E-mail: cosmai@@bari.infn.it}

\begin{abstract} 
We analyze the problem of the Nielsen-Olesen unstable modes in
the SU(2) lattice gauge theory by means of a recently introduced
gauge-invariant effective action. We perform numerical simulations
in the case of a constant Abelian chromomagnetic field. We find
that for lattice sizes above a certain critical length the density
of effective action shows a behaviour compatible with the presence
of the unstable modes.
We put out a possible relation between the dynamics of the unstable 
modes and the confinement.
\end{abstract}
\end{frontmatter}

\section{Introduction}

It is well known since long time that for non-Abelian gauge theories
without matter fields in the one-loop approximation states with an
external constant Abelian chromomagnetic field lies below the
perturbative ground states. In the case of the SU(2) gauge theory the
vacuum energy density turns out to be~\cite{Savvidy77} in the one-loop
approximation
\begin{equation}
\label{energy}
 \varepsilon(H)  = \frac{1}{2} H^2 + \frac{11}{48 \pi^2} g^2H^2
 \ln(\frac{gH}{\Lambda^2}) + {\mathcal O}(g^2H^2)  \,,
\end{equation}
where $H$ is the strength of the external magnetic field. The energy
density Eq.(\ref{energy}) has a negative minimum at some non zero value of the
external field $H$. However, N. K. Nielsen and P.
Olesen~\cite{Nielsen78} pointed out that the Savvidy states are not
stable due to long-range modes: the Nielsen-Olesen unstable modes.

In a series of papers the proposal was advanced that the stabilization of the
Nielsen-Olesen unstable modes results in the formation of a quantum liquid, the
so-called Copenhagen vacuum~\cite{Nielsen81}.

The problem of the Nielsen-Olesen unstable modes has been reconsidered
by using variational techniques on a class of gauge-invariant Gaussian
wave functionals~\cite{Consoli85,Cea87}. It turned out that the
unstable modes contribute to the vacuum energy density with a negative
classical term. In particular, in Ref.~\cite{Cea87} it was shown that
the Nielsen-Olesen modes get stabilized by considering configurations
which differ from the classical external field only in the unstable
sector. Moreover, the configurations which minimize the energy density
screen almost completely the classical external chromomagnetic field
and contribute to the energy density with a negative classical term
that cancels the positive classical term in Eq.({\ref{energy}).
However, it was pointed out that the calculation of the one-loop
contribution to the vacuum energy density is  truly non-perturbative
and needs a completely non-perturbative approach.

Several authors studied gauge theories with an external background
field on the lattice, both in three~\cite{Cea93,Trottier93} and
four~\cite{Ambjorn90,Levi93} space-time dimensions. In
particular, the results of Refs.~\cite{Cea93}
and~\cite{Levi93} showed some evidence indicating the presence of
the unstable modes. In those papers, however, the external background field was 
introduced via an external source. As a consequence the lattice
effective action loses the manifest gauge invariance, for the source
term in the lattice action becomes invariant only under a  subgroup of
the gauge group. 

The aim of the present paper is to analyze the problem of the 
unstable modes in the case of the SU(2) lattice gauge theory by means
of the recently introduced gauge invariant effective
action~\cite{Cea95}.

\section{SU(2) in a Constant Background Field}

Let us consider a static external background field
$\vec{A}_a^{\text{ext}}(\vec{x})$. We define the effective action which is
invariant against gauge transformation of the external field by using the
so-called Schr\"odinger functional:
\begin{equation}
\label{Z}
{\mathcal{Z}} \left[ \vec{A}_a^{\text{ext}} \right] =
 \langle  \vec{A}_a^{\text{ext}}  |  \exp(-HT) {\mathcal{P}} |
\vec{A}_a^{\text{ext}}
\rangle  \,,
\end{equation}
where ${\mathcal{P}}$ projects onto the physical states. On the
lattice we can rewrite ${\mathcal{Z}} [ \vec{A}_a^{\text{ext}} ]$ as
follows
\begin{equation}
\label{Zlatt}
{\mathcal{Z}} \left[ \vec{A}_a^{\text{ext}} \right] =
\int {\mathcal{D}}U \exp(-S_W) \,,
\end{equation}
where $S_W$ is the standard Wilson action, and we integrate over the
lattice links $U_\mu(x)$ with the constraints
\begin{equation}
\label{constraints}
U_\mu(x)|_{x_4=0} = U_\mu^{\text{ext}}(\vec{x},0) \,.
\end{equation}
The external links are related to the continuum gauge field
${\vec{A}}^{\text{ext}}_a$ by the well-known relation
\begin{equation}
\label{Umu}
U_\mu^{\text{ext}}(x) = {\text{P}} \exp \left\{ + iag  \int_0^1 dt \,
A_\mu^{\text{ext}}(x+ at {\hat{\mu}}) \right\}
\end{equation}
with 
$A_0^{\text{ext}}(\vec{x}) =0$ and 
$\vec{A}^{\text{ext}}(\vec{x}) =  \vec{A}_a^{\text{ext}}(\vec{x})
\lambda_a/2$, and ${\text{P}}$ is the path-ordering operator. It
should be emphasized that the lattice Schr\"odinger functional has a
well-defined continuum limit and, due to the gauge-invariance, does
not require extra counterterms~\cite{Luscher92}. The lattice effective
action for the external background field
$\vec{A}_a^{\text{ext}}(\vec{x})$ is given by
\begin{equation}
\label{Gamma}
\Gamma\left[ \vec{A}^{\text{ext}} \right] = -\frac{1}{T}
\ln \left\{ \frac{{\mathcal{Z}}[\vec{A}^{\text{ext}}]}{{\mathcal{Z}}(0)} \right\}
\end{equation}
where $T$ is the extension in the Euclidean time. It is
straightforward to show that, in the limit $T \rightarrow \infty$, 
$\Gamma[ \vec{A}^{\text{ext}}]$ reduces to the vacuum energy in
presence of the given background field. Thus, our effective action is
relevant to investigate non perturbatively the properties of the
quantum vacuum. 

In this paper we are interested in the case of a
constant abelian chromomagnetic field for the SU(2) gauge theory
\begin{equation}
\label{field}
\vec{A}_a^{\text{ext}}(\vec{x}) =\vec{A}^{\text{ext}}(\vec{x})\delta_{a,3} 
\,, \,\,\,
A_k^{\text{ext}}(\vec{x}) = \delta_{k,2} x_1 H \,.
\end{equation}
Note that, due to the gauge invariance, 
$\Gamma[\vec{A}^{\text{ext}}]$ depends on the constant field strength
$F^3_{12}=H$. Thus, the relevant quantity is the density of the
effective action:
\begin{equation}
\label{density}
\varepsilon\left[ \vec{A}^{\text{ext}} \right] =
-\frac{1}{\Omega} \ln \left[
\frac{{\mathcal{Z}}[ \vec{A}^{\text{ext}}]}{{\mathcal{Z}}(0)} \right]
\,, 
\end{equation}
where $\Omega = V \cdot T$, and $V$ is the spatial volume. The
external links corresponding to $\vec{A}^{\text{ext}}_a$ are easily
evaluated from Eq.(\ref{Umu}). We get
\begin{eqnarray}
\label{Umuext}
&&  U^{\text{ext}}_2(\vec{x},0) = \cos\left(\frac{agHx_1}{2}\right)  + i \sigma^3
 \sin\left(\frac{agHx_1}{2}\right)
\nonumber \\
&&  U^{\text{ext}}_1(\vec{x},0) =  U^{\text{ext}}_3(x) = U^{\text{ext}}_4(x) = 1
\,.
\end{eqnarray}
By imposing the periodic boundary conditions we obtain the
quantization condition
\begin{equation}
\label{quantz}
\frac{a^2 g H}{2} = \frac{2 \pi}{L_1} n^{\text{ext}}
\end{equation}
where $n^{\text{ext}}$ is an integer,  $L_1$ the lattice extension in the
$x_1$ direction (in lattice units).

\section{The Unstable Modes}

Let us now briefly discuss the origin of the unstable modes both in
the continuum and on the lattice. 

In order to evaluate the one-loop effective action in the continuum
one writes
\begin{equation}
\label{bf}
A^a_\mu(x) =  \bar{A}^a_\mu(x) + \eta^a_\mu(x)
\end{equation}
where $\bar{A}^a_\mu(x)=\delta_{\mu 2} \delta^{a 3} x_1 H$ and 
$\eta(x)$ is the quantum fluctuation on the background field.
In the background gauge 
\begin{equation}
\label{backg}
\left( \delta^{ab} \partial_\mu - g \varepsilon^{a b c} \bar{A}^c_\mu
\right) \eta^b_\mu(x) = 0 \,,
\end{equation}
we rewrite the pure gauge action in the one-loop approximation as
\begin{equation}
\label{SYM}
S_{Y-M}= S_{\text{class}} + \frac{1}{2} \int d^4x \, \eta^a_\mu(x) 
{\mathcal{O}}^{ab}_{\mu \nu} \eta^b_\nu(x) \,.
\end{equation}
The one-loop effective action can be obtained by performing the
Gaussian functional integrations over the quantum fluctuations and
including the Faddeev-Popov one-loop determinant. However, if we solve
the eigenvalue equations
\begin{equation}
\label{eigen}
{\mathcal{O}}^{ab}_{\mu \nu} \phi^b_\nu(x) = \lambda \phi^a_\mu(x) \,,
\end{equation}
then we find that there are negative eigenvalues:
\begin{equation}
\label{negeigens}
\lambda_u = p_0^2 + p_3^2 - gH  \,.
\end{equation}
Indeed $\lambda_u <0$ when $gH >  p_0^2 + p_3^2 $. The modes with
eigenvalue $\lambda_u$ are the Nielsen-Olesen unstable modes. If we
perform formally the Gaussian functional integration, then the
one-loop effective action picks out an imaginary part. The point is
that in the functional integration over the unstable modes one must
include the positive quartic term. If we do that, we find that the
unstable modes behave like a two-dimensional tachyonic charged scalar
field. Thus the stabilization of the unstable modes resemble the
(dynamical) Higgs mechanism. The negative condensation energy we gain
in the stabilization procedure cancels the positive classical magnetic
energy~\cite{Cea87}.

On the lattice we would like to evaluate the Schr\"odinger functional
Eq.(\ref{Zlatt}) in the weak coupling region. To this end, we
write~\cite{Luscher92}
the lattice version of Eq.(\ref{bf}):
\begin{equation}
\label{bflatt}
U_\mu(x) = \exp \left( iagq_\mu(x) \right ) U_\mu^{\text{ext}}(x) \,,
\end{equation}
where the fluctuations $q_\mu(x) = q^a_\mu(x) \sigma^a/2$ satisfy the
boundary conditions
\begin{equation}
\label{qmu}
q_\mu(x) |_{x_4=0} = 0\,.
\end{equation}
Inserting Eq.(\ref{bflatt}) into the Wilson action we obtain in the
one-loop approximation 
\begin{equation}
S_W = S^{\text{ext}} + S^{(2)}
\end{equation}
where 
\begin{equation}
S^{\text{ext}} = \frac{4 \Omega}{g^2} \left[ 1 - \cos \left(
\frac{gH}{2} \right) \right]
\end{equation}
and $S^{(2)}$ is quadratic in the quantum fluctuations $q_\mu(x)$. We are
interested in the spectrum of the lattice version of the operator
${\mathcal{O}}^{ab}_{\mu\nu}$ in Eq.(\ref{eigen}). Unfortunately the
discrete eigenvalue equation cannot be solved (analytically) in closed
form. Obviously it is possible to solve the eigenvalue equation by means of
numerical methods. We plan to discuss this matter in a future paper.
Nevertheless, if we discard a lot of irrelevant operators in
$S^{(2)}$, i.e. terms which vanish in the naive continuum limit, we
find an approximate formula for the eigenvalues. In particular the
unstable mode eigenvalue turns out to be
\begin{equation}
\label{umode}
\lambda_u = (1 -\cos p_4) + (1 -\cos p_3) - \sin \left( \frac{gH}{2}
\right) \,.
\end{equation}
By using Eq.(\ref{quantz}) and $p_\mu= 2\pi  n_\mu/L_\mu$, we get 
\begin{equation}
\label{lambdau}
\lambda_u =  1- \cos \left( \frac{2 \pi}{L_4} n_4 \right) +  1 -
\cos \left( \frac{2 \pi}{L_3} n_3 \right) -
\sin \left( \frac{2 \pi}{L_1} n^{\text{ext}}   \right) \,.
\end{equation}

\section{Monte Carlo Simulations}

We performed our Monte Carlo simulations on lattices with linear sizes
$L_1 = L_4 = 32$, $L_2 = L_3 = L$ and $n^{\text{ext}}=1$. Inserting
these values into Eq.(\ref{lambdau}), it follows that $\lambda_u < 0$
when $L \ge L_{\text{crit}}$, $L_{\text{crit}} \simeq  10$.
Remarkably, these considerations suggest that we can switch on and off
the unstable modes by varying $L$. 

We turn now on discussing the outcomes of the numerical simulations.
We are interested in the vacuum energy density Eq.({\ref{density}). To
avoid the problem of computing a partition function, we focus on the
derivative of $\varepsilon[\vec{A}^{\text{ext}}]$ with respect to
$\beta$ by taking $n^{\text{ext}}$ (i.e. $gH$, see Eq.(\ref{quantz})) fixed. A
straightforward calculation gives:
\begin{equation}
\label{epsilonp}
 \varepsilon^{\prime} \left[ \vec{A}^{\text{ext}} \right] =
\left \langle \frac{1}{\Omega}
 \sum_{x,\mu>\nu} \frac{1}{2} \text{Tr}
 U_{\mu\nu}(x) \right \rangle_0 \, -
 \left \langle \frac{1}{\Omega} \sum_{x,\mu>\nu} \frac{1}{2} \text{Tr}
U_{\mu\nu}(x) \right \rangle_{A^{\text{ext}}} 
\end{equation}
where the $U_{\mu\nu}(x)$'s are the plaquettes in the $(\mu,\nu)$
plane. As we said, our simulations are performed on $32 \times L^2
\times 32$ lattices with $L=6,8,10,16,20,24$. We impose periodic
boundary conditions. We use standard overrelaxed Metropolis algorithm
to update gauge configurations. The links belonging to the time slice
$x_4=0$ are frozen according to Eq.(\ref{Umuext}). Moreover, we also
impose that the links at the spatial boundaries are fixed according to
Eq.(\ref{Umuext}). Note that this condition corresponds in the
continuum to the usual requirement that the fluctuations over the 
background field vanish at the
spatial infinity. 
It is obvious that the contributions to  $\varepsilon^{\prime} [
\vec{A}^{\text{ext}} ]$  due to the frozen time slice at
$x_4=0$ and to the fixed links at the lattice spatial
boundaries must be subtracted, i.e.
only the dynamical
links must be taken into account in evaluating
$\varepsilon^{\prime} [\vec{A}^{\text{ext}} ]$.
We denote by  $\Omega=L_1 L_2 L_3 L_4$ the total number of lattice
sites (i.e. the lattice volume) belonging to the lattice $\Lambda$.
$\Omega_{\text{ext}}$ are the lattice sites whose links
are fixed according to Eq.(\ref{Umuext}):
\begin{equation}
\label{Omegaext}
\Omega_{\text{ext}}   =    L_1 L_2 L_3 +  (L_4-1)
\times  (L_1 L_2 L_3 - (L_1-2)(L_2-2)(L_3-2)) \,.
\end{equation}
Hence $\Omega_{\text{int}}=\Omega-\Omega_{\text{ext}}$ is the volume occupied
by the  ``internal'' lattice sites (let us denote this ensemble by
$\tilde{\Lambda}$). Accordingly we define the derivative of the ``internal''
energy density $\varepsilon^{\prime}_{\text{int}}$ as
\begin{equation}
\label{epsilonpint}
 \varepsilon^{\prime}_{\text{int}} \left[ \vec{A}^{\text{ext}} \right] =
\left \langle \frac{1}{\Omega_{\text{int}}}
 \sum_{x \in \tilde{\Lambda},\mu>\nu} \frac{1}{2} \text{Tr}
 U_{\mu\nu}(x) \right \rangle_0 \, -
 \left \langle \frac{1}{\Omega_{\text{int}}} 
\sum_{x \in \tilde{\Lambda},\mu>\nu} \frac{1}{2} \text{Tr}
U_{\mu\nu}(x) \right \rangle_{A^{\text{ext}}} \,.
\end{equation}
In Figures 1 and 2 we report the derivative  of the internal energy density
versus $\beta$ in units of the derivative of the energy density due to the
``external'' links defined in Eq.(\ref{Umuext})
\begin{equation}
\label{epspext}
\varepsilon^{\prime}_{\text{ext}} = \frac{\partial}{\partial \beta}
\frac{1}{\Omega} S^{\text{ext}} = 1- \cos \left( \frac{2 \pi}{L_1}
n^{\text{ext}} \right) \,.
\end{equation}
Several features of Fig.~1 are worth noting. In the strong coupling
region $\beta \lesssim 1$ the external background field is completely
shielded. This behavior is very similar to the one we found for the
U(1) gauge theory~\cite{Cea95}. Moreover,
$\varepsilon^{\prime}_{\text{int}}$  displays a peak at $\beta \simeq
2.2$. 
These features are displayed
also~\cite{Lautrup80} by the specific heat in pure SU(2). This is not
surprising, for our previous studies~\cite{Cea95,Cea96} in U(1) showed
that  $\varepsilon^{\prime}_{\text{int}}$ behaves as the specific
heat. 

The derivative of the internal energy density displays the most
interesting behavior in the weak coupling region $\beta \gtrsim 2.3$.
Indeed, we feel that this peculiar behavior is due to the unstable
modes. To see this, we recall that according to our previous
discussion we have in the weak coupling region
\begin{equation}
\label{epsbnext}
\varepsilon_{\text{int}}(\beta, n^{\text{ext}}) = \beta \left[ 1 - \cos \left(
\frac{2 \pi}{L_1} n^{\text{ext}} \right) \right] + g(n^{\text{ext}}) +
{\mathcal{O}} \left( \frac{1}{\beta} \right) \,.
\end{equation}
The first term in Eq.(\ref{epsbnext}) is the classical magnetic
energy, while $g(n^{\text{ext}})$ is the one-loop contribution. As a
consequence we expect that
\begin{equation}
\label{epsbnext-prime}
\varepsilon_{\text{int}}^{\prime}(\beta, n^{\text{ext}}) = 
1 - \cos \left( \frac{2 \pi}{L_1} n^{\text{ext}} \right) 
+ {\mathcal{O}} \left( \frac{1}{\beta^2} \right) \,.
\end{equation}
Equation~({\ref{epsbnext-prime}) tells us that, in the weak coupling
region the derivative of the vacuum energy density reduces to 
$\varepsilon^{\prime}_{\text{ext}}$. Indeed, in the U(1) gauge theory
this is the case~\cite{Cea95}. From Figure~1 we see that , for $L \simeq
6-8$,  $\varepsilon^{\prime}_{\text{int}}$ tends to 
$\varepsilon^{\prime}_{\text{ext}}$ (in the weak coupling region) as in U(1). 
However, by increasing
$L$ the ratio 
$\varepsilon^{\prime}_{\text{int}}/\varepsilon^{\prime}_{\text{ext}}$
decreases (see Fig.~2).

In order to ascertain if the effects displayed by our data are not due to the
lattice artifacts, we contrast in Fig.3 the derivative of the vacuum energy
density in the weak coupling region both for SU(2) and U(1) versus the lattice
linear size $L$.  As we can see, the ratio 
$\varepsilon^{\prime}_{\text{int}}/\varepsilon^{\prime}_{\text{ext}} \simeq 1$
and it is almost independent on $L$ in the case of the $U(1)$ theory. On the
other hand, for the SU(2) case Fig.3 clearly shows that the ratio
$\varepsilon^{\prime}_{\text{int}}/\varepsilon^{\prime}_{\text{ext}}
\rightarrow 0$  in the infinite volume limit. As we shall argue in a moment,
this peculiar behaviour can be ascribed to the unstable modes. To see this we
recall that the stabilization of the unstable modes should modify
Eq.(\ref{epsbnext}) as follows~\cite{Cea87}:
\begin{equation}
\label{epsbnext-new}
\varepsilon_{\text{int}}(\beta, n^{\text{ext}}) = 
g(n^{\text{ext}}) + {\mathcal{O}} \left( \frac{1}{\beta} \right) \,.
\end{equation}
However, in Ref.~\cite{Cea87} it was pointed out that the cancellation
of the classical energy term happens only in the thermodynamical limit
$V \rightarrow \infty$. So that we expect that
Eq.(\ref{epsbnext-new}) holds on the infinite lattice.
On a finite lattice we expect that in the weak coupling region 
\begin{equation}
\label{ratio}
\frac{\varepsilon^{\prime}_{\text{int}}}{\varepsilon^{\prime}_{\text{ext}}} = 
f(x) + {\mathcal{O}} \left( \frac{1}{\beta^2} \right)
\end{equation}
with $x=a_H/L_{\text{eff}}$, where $a_H = \sqrt{2 \pi/gH}$ is the 
magnetic length and
\begin{equation}
\label{Leff}
L_{\text{eff}} = \Omega_{\text{int}}^{1/4} \,,
\end{equation}
is the lattice effective linear size.
The continuum calculations of Ref.~\cite{Cea87} tell us that $f(0)=0$. In Figure~4 we display the
ratio
$\varepsilon^{\prime}_{\text{int}}/\varepsilon^{\prime}_{\text{ext}}$ for $\beta=5$ versus the
lattice effective linear size. Indeed Fig.4 shows   
that $\varepsilon^{\prime}_{\text{int}}$ decreases by increasing $L_{\text{eff}}$.
So that we have 
$\varepsilon^{\prime}_{\text{int}} \rightarrow 0$ when
$L_{\text{eff}} \rightarrow \infty$, in agreement with
Eq.(\ref{epsbnext-new}). Moreover, we see clearly that a critical
length exists such that for $L_{\text{eff}} \le
L^{\text{crit}}_{\text{eff}}$ Eq.(\ref{epsbnext-prime}) holds. It is
remarkable that our extimation of the critical length 
\begin{equation}
\label{critlength}
L^{\text{crit}}_{\text{eff}} \approx 12
\end{equation}
is in good agreement with Ref.~\cite{Levi93}.
Moreover, in the region $L_{\text{eff}} > L^{\text{crit}}_{\text{eff}}$ we fitted the data trying
the power law
\begin{equation}
\label{powerlaw}
f(x) = k x^\alpha \,.
\end{equation}
Indeed we find a rather good fit ($\chi^2/{\text{d.o.f.}} \simeq 0.3$, solid
line in Fig.4) with
\begin{equation}
\label{kappaalpha}
k = 5 (1) \quad , \qquad \alpha  =  1.43 (8)   \,.
\end{equation}
In our opinion the existence of a critical length and the cancellation of the
classical magnetic energy density in the thermodynamical limit constitute a
rather strong evidence for the unstable modes on the lattice.  Remarkably it
turns out that also the peak value of  $\varepsilon^{\prime}_{\text{int}}$
tends towards zero with the same power law as implied by
Eq.~(\ref{kappaalpha}). Indeed, in Fig.~5 we show the ratio
$\varepsilon^{\prime}_{\text{int}}/\varepsilon^{\prime}_{\text{ext}}$  taken at
the peaks in Figs.~1 and 2. For   $L_{\text{eff}} >
L^{\text{crit}}_{\text{eff}}$  we find (solid line in Fig.5)
\begin{equation}
\label{ratio1}
\frac{\varepsilon^{\prime}_{\text{int}}}{\varepsilon^{\prime}_{\text{ext}}} = 
k^{\prime} \left( \frac{a_H}{L_{\text{eff}}} \right)^{\alpha^{\prime}} \,,
\end{equation}
where $k^{\prime} = 10(2)$, $\alpha^{\prime} = 1.5 (1)$ and 
$\chi^2/{\text{d.o.f.}} \simeq 0.3$. Note that $\alpha^{\prime} = \alpha$
within our statistical uncertainties. Thus Eqs.~(\ref{ratio}),
(\ref{powerlaw}), and~(\ref{ratio1}) imply that 
$\varepsilon^{\prime}_{\text{int}}[\vec{A}^{\text{ext}}]$ tends uniformly
towards zero in the thermodynamical limit.

\section{Conclusions}

Let us conclude by stressing the main results of this paper.

By using the gauge invariant effective action for background fields on the
lattice we studied non perturbatively the dynamics of the Nielsen-Olesen
unstable modes. We feel that the existence of a critical length
$L^{\text{crit}}_{\text{eff}}$ and the cancellation of the classical magnetic
energy in the infinite volume limit constitute a strong evidence for the
unstable modes on the lattice. Moreover, 
our numerical results are suggesting that
$\varepsilon^{\prime} \left[ \vec{A}^{\text{ext}} \right]
\rightarrow 0$ when $L_{\text{eff}} \rightarrow \infty$ in the whole
$\beta$-region. Thus in the continuum limit
$L_{\text{eff}} \rightarrow \infty$, $\beta \rightarrow \infty$
the confining vacuum screens completely the external chromomagnetic Abelian
field. In other words, the continuum vacuum behaves as an Abelian
 magnetic condensate medium in accordance with the dual superconductivity
scenario. The magnetic condensate dynamics seems to be closely
related to the presence of the Nielsen-Olesen unstable modes.
In conclusion we feel that our approach to a gauge-invariant 
effective action on the
lattice is a useful tool to understand the dynamics of
color confinement.

\clearpage

\section*{FIGURE CAPTIONS}

\renewcommand{\labelenumi}{Figure \arabic{enumi}.}
\begin{enumerate}
\item
$\varepsilon^{\prime}_{\text{int}}[ \vec{A}^{\text{ext}} ]/
 \varepsilon^{\prime}_{\text{ext}}$ versus $\beta$
for $L=6$ (stars) and $L=8$ (crosses).
\item
$\varepsilon^{\prime}_{\text{int}}[ \vec{A}^{\text{ext}} ]/
 \varepsilon^{\prime}_{\text{ext}}$
versus $\beta$
for $L=10$ (diamonds), $L=16$ (triangles) and $L=24$ (circles).
\item
The ratio 
$\varepsilon^{\prime}_{\text{int}}[ \vec{A}^{\text{ext}} ]/
 \varepsilon^{\prime}_{\text{ext}}$
in the weak coupling region
 versus $L$ for $U(1)$ (circles, $\beta=3$) and $SU(2)$ (squares, $\beta=5$).
\item
The ratio
$\varepsilon^{\prime}_{\text{int}}[ \vec{A}^{\text{ext}} ]/
 \varepsilon^{\prime}_{\text{ext}}$ at $\beta=5$ versus $L_{\text{eff}}$.
The solid line is the fit Eq.~(\ref{powerlaw}).
\item
The ratio
$\varepsilon^{\prime}_{\text{int}}[ \vec{A}^{\text{ext}} ]/
 \varepsilon^{\prime}_{\text{ext}}$ 
at the peak values versus $L_{\text{eff}}$.
The solid line is the fit Eq.~(\ref{ratio1}).
\end{enumerate}

\newpage
\begin{figure}[t]
\label{Fig1}
\begin{center}
\epsfig{file=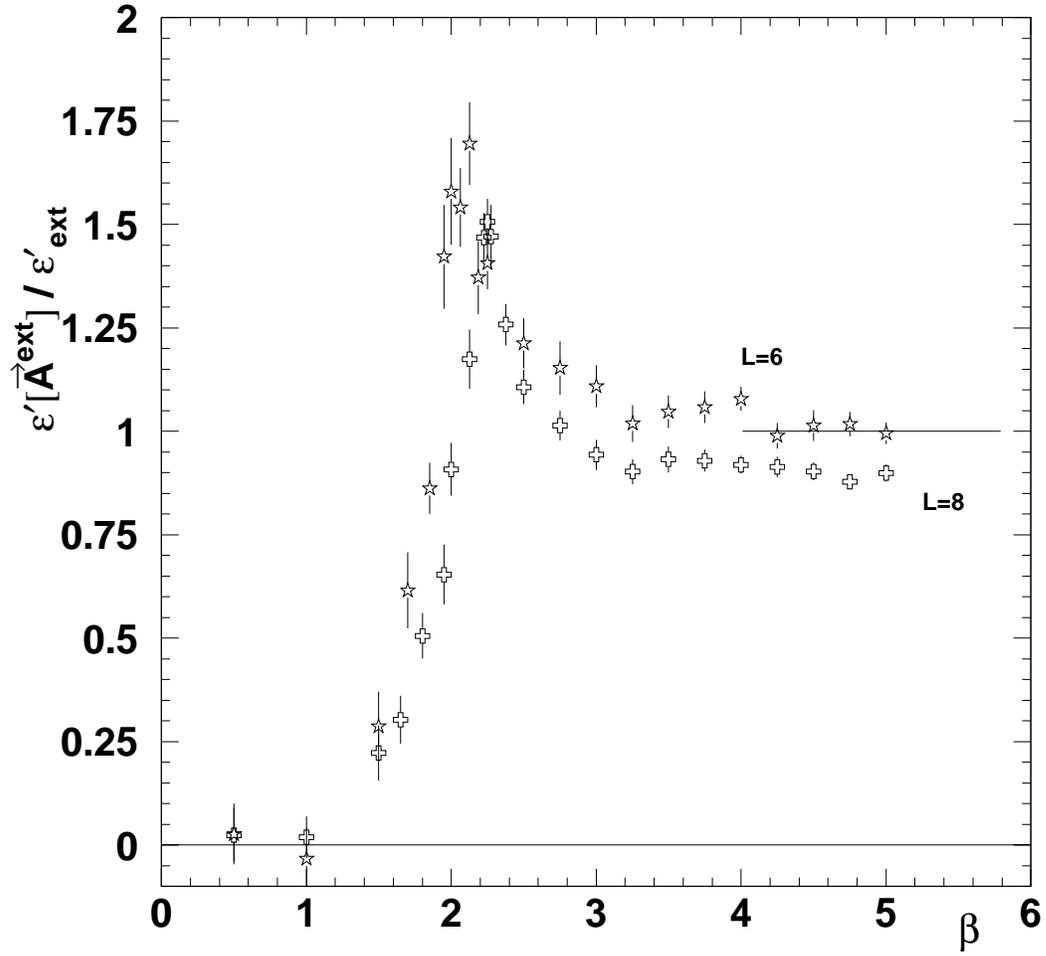,width=\textwidth}
\caption{ $\varepsilon^{\prime}_{\text{int}}[ \vec{A}^{\text{ext}} ]/
 \varepsilon^{\prime}_{\text{ext}}$ versus $\beta$
for $L=6$ (stars) and $L=8$ (crosses).}
\end{center}
\end{figure}
\newpage
\begin{figure}[t]
\begin{center}
\label{Fig2}
\epsfig{file=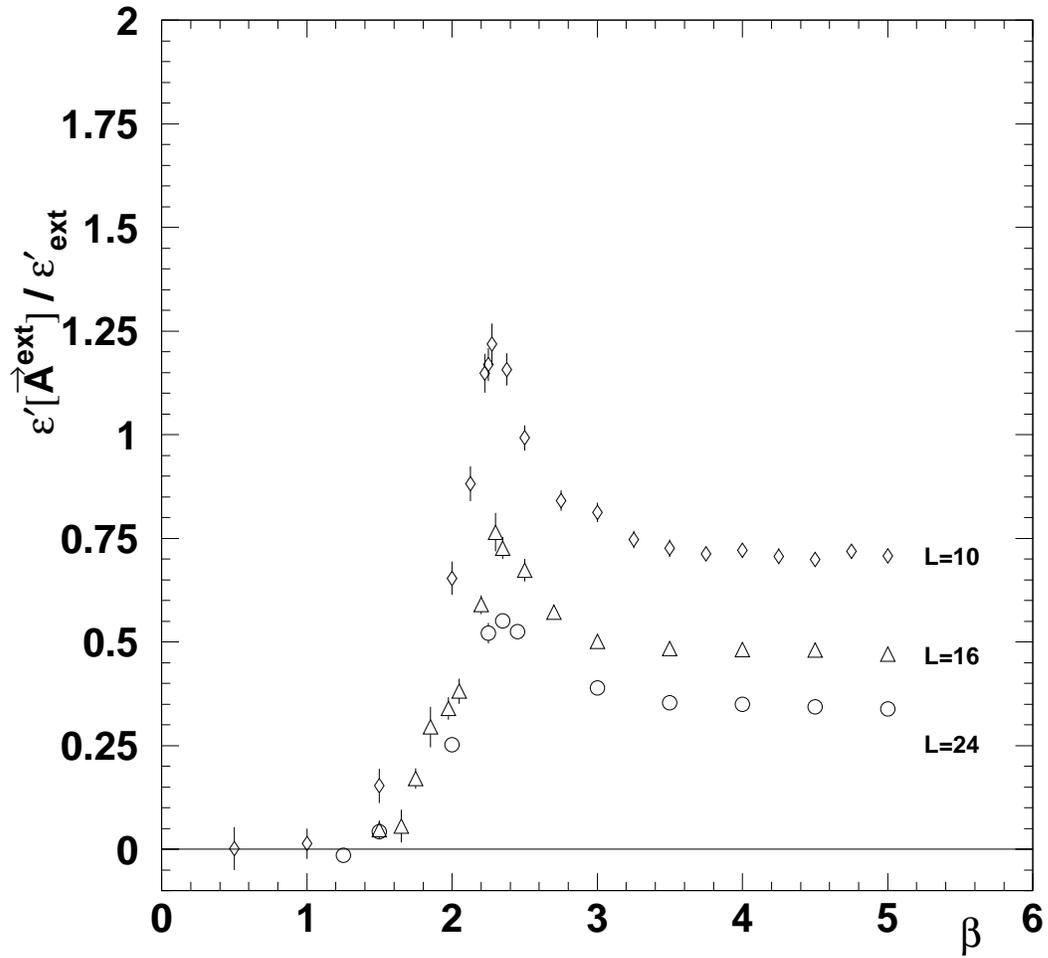,width=\textwidth}
\caption{ $\varepsilon^{\prime}_{\text{int}}[ \vec{A}^{\text{ext}} ]/
 \varepsilon^{\prime}_{\text{ext}}$
versus $\beta$
for $L=10$ (diamonds), $L=16$ (triangles) and $L=24$ (circles).}
\end{center}
\end{figure}
\newpage
\begin{figure}[t]
\label{Fig3}
\begin{center}
\epsfig{file=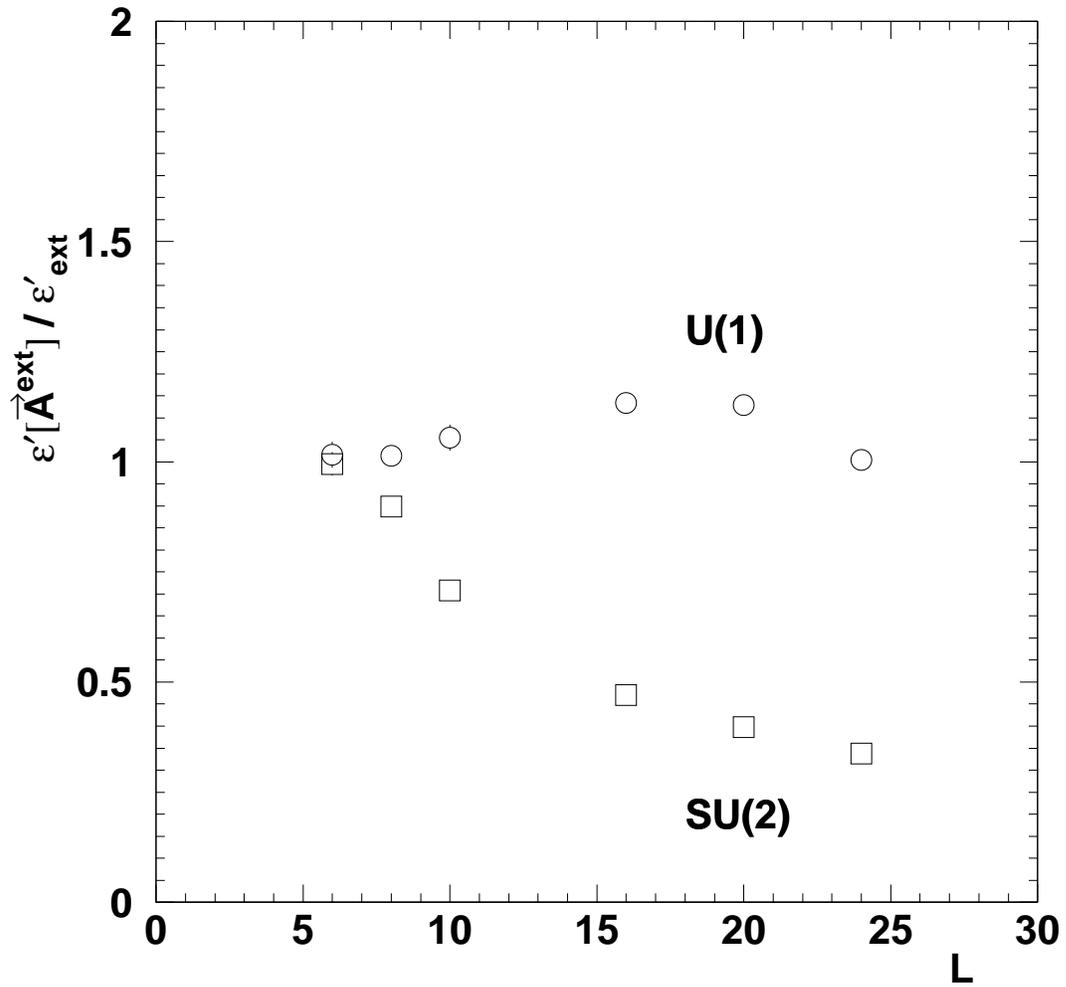,width=\textwidth}
\caption{The ratio 
$\varepsilon^{\prime}_{\text{int}}[ \vec{A}^{\text{ext}} ]/
 \varepsilon^{\prime}_{\text{ext}}$
in the weak coupling region
 versus $L$ for $U(1)$ (circles, $\beta=3$) and $SU(2)$ (squares, $\beta=5$).}
\end{center}
\end{figure}
\newpage
\begin{figure}[t]
\label{Fig4}
\begin{center}
\epsfig{file=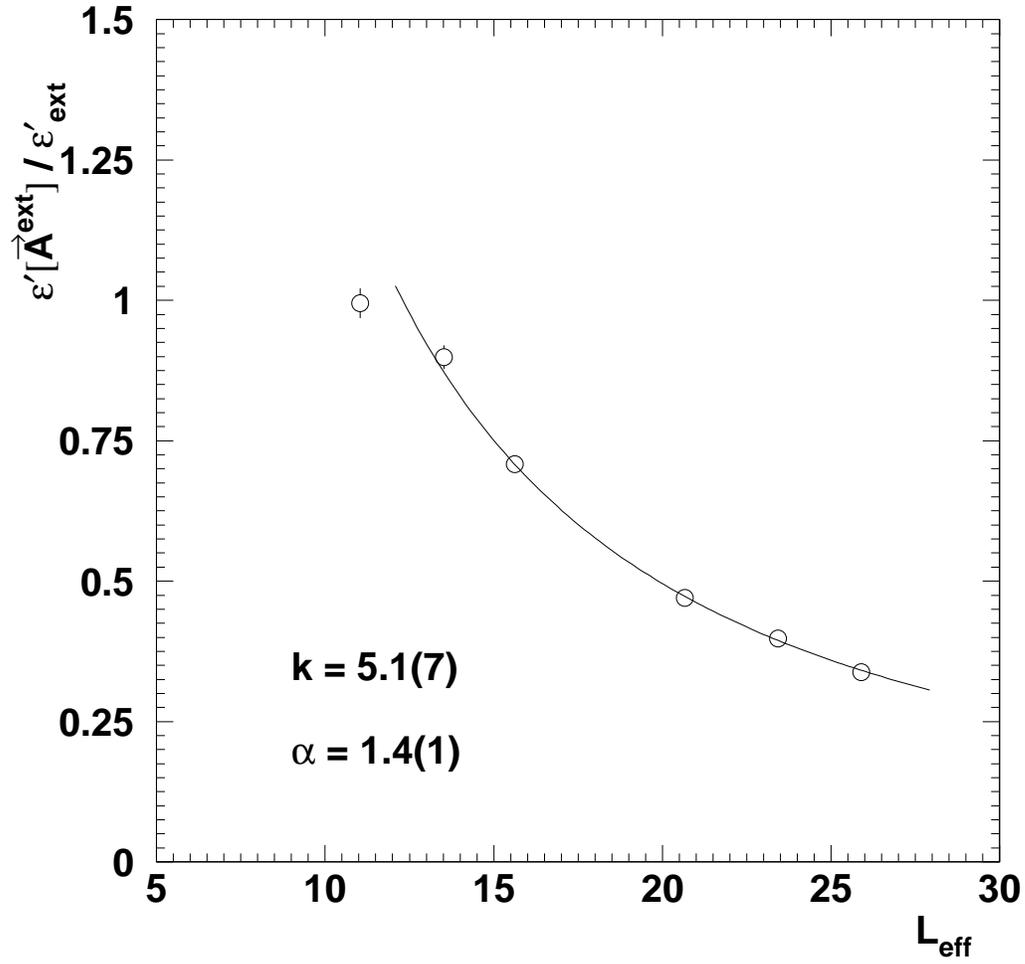,width=\textwidth}
\caption{The ratio
$\varepsilon^{\prime}_{\text{int}}[ \vec{A}^{\text{ext}} ]/
 \varepsilon^{\prime}_{\text{ext}}$ at $\beta=5$ versus $L_{\text{eff}}$.
The solid line is the fit Eq.~(\ref{powerlaw}).
}
\end{center}
\end{figure}
\newpage
\begin{figure}[t]
\label{Fig5}
\begin{center}
\epsfig{file=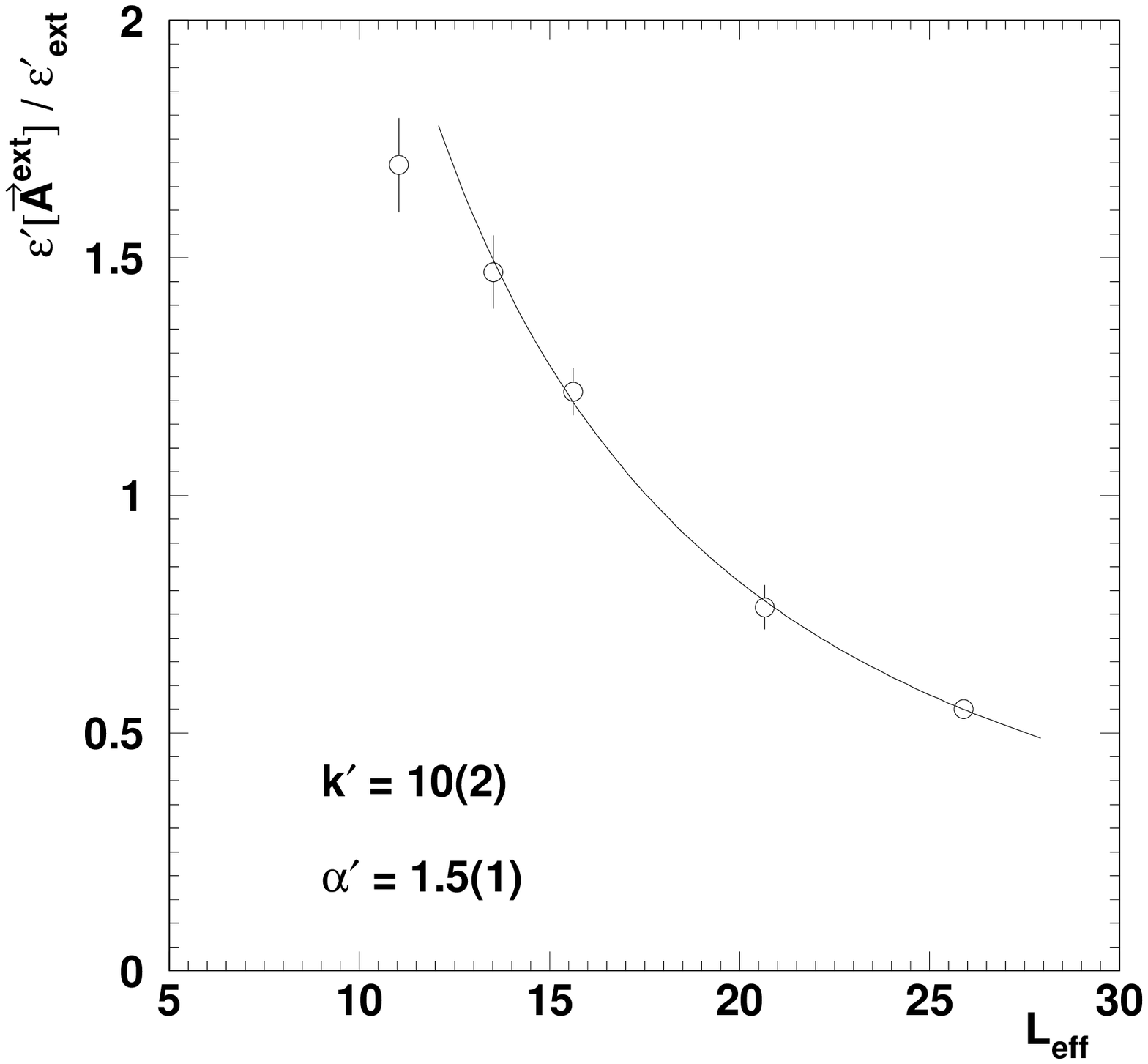,width=\textwidth}
\caption{The ratio
$\varepsilon^{\prime}_{\text{int}}[ \vec{A}^{\text{ext}} ]/
 \varepsilon^{\prime}_{\text{ext}}$ 
at the peak values versus $L_{\text{eff}}$.
The solid line is the fit Eq.~(\ref{ratio1}).
}
\end{center}
\end{figure}


\begin{thebibliography}{99}
\bibitem{Savvidy77} G. K. Savvidy, Phys. Lett. {\bf B71} (1977) 133;
S. G. Matinyan and G. K. Savvidy, Nucl. Phys. {\bf B134} (1978) 539;
I. A. Balatin, S. G. Matinyan   and G. K. Savvidy, Sov. J. Nucl. Phys.
{\bf 26} (1978) 214.
\bibitem{Nielsen78} N.K. Nielsen and P. Olesen, Nucl. Phys. {\bf B144}
 (1978) 376.
\bibitem{Nielsen81} For a review, see: H. B. Nielsen in {\em Particle
Physics}, eds. I Andri\'c, I. Dadi\'c and N. Zovko (North-Holland,
Amsterdam, 1981).
\bibitem{Consoli85} M. Consoli and G. Preparata,  Phys. Lett. {\bf 154B}
(1985) 411.
\bibitem{Cea87} P. Cea, Phys. Rev.  {\bf D37} (1988) 1637.
\bibitem{Cea93} P. Cea and L. Cosmai,  Phys. Rev. {\bf D48} (1993)
3364; P. Cea and L. Cosmai,  hep-lat/9306007; P. Cea and L. Cosmai, Nucl. Phys. (Proc. Suppl.)
{\bf 34} (1994) 234.
\bibitem{Trottier93} H. D. Trottier and  R. M. Woloshyn,
Phys. Rev. Lett. {\bf 70} (1993) 2053; E.~{\bf 72} (1994) 4155.
\bibitem{Ambjorn90} J. Ambjorn, V. K. Mitriushkin, and A. M.
Zadorozhny, Phys. Lett. {\bf B245} (199) 575; 
P. Cea and L. Cosmai, Phys. Lett. {\bf B264} (1991) 415.
\bibitem{Levi93} A. R. Levi, Nucl. Phys. B (Proc. Suppl.) {\bf 34}
(1994) 161; A. R. Levi and J. Polonyi,  Phys. Lett. {\bf B357} (1995)
186.
\bibitem{Cea95} P. Cea, L. Cosmai, and A. D. Polosa, hep-lat/9601010.
\bibitem{Luscher92} M. L\"uscher, R. Narayanan, P. Weisz, and U.
Wolff, Nucl. Phys. {\bf B384} (1992) 168; 
M. L\"uscher and   P. Weisz,   Nucl. Phys. {\bf B452} (1995) 213.
\bibitem{Lautrup80} B. Lautrup and M. Nauenberg, Phys. Rev. Lett. {\bf
45} (1980) 1755; J. Engels and T. Scheideler, hep-lat/960741.
\bibitem{Cea96} P. Cea, L. Cosmai, and A. D. Polosa, hep-lat/9607020.


\end{thebibliography}
\end{document}